\documentclass[aps,prb,twocolumn,showpacs,superscriptaddress]{revtex4-1}
\usepackage{amssymb}
\usepackage{amsmath}
\usepackage{amsthm}
\usepackage{balance}

\usepackage{graphicx}
\usepackage{amsfonts}
\usepackage{bm}  
\usepackage{natbib} 
\usepackage{hyperref}
\usepackage{color} 

\usepackage{esint} 
\usepackage{siunitx} 
\definecolor{midnightblue}{cmyk}{1,1,0,0.1}
\definecolor{forestgreen}{cmyk}{0.76,0,0.26,0.5} 
\definecolor{cred}{RGB}{188,55,84} 
\hypersetup{colorlinks=true,linkcolor=cred,citecolor=cred,urlcolor=cred}

\def\be{\begin{equation}} \def\ee{\end{equation}}
\def\bea{\begin{eqnarray}} \def\eea{\end{eqnarray}}

\def\fracp#1#2{\frac{\partial #1}{\partial #2}}

\newcommand{\ket}[1]{| #1 \rangle}
\newcommand{\bra}[1]{\langle #1 |}
\newcommand{\Lagr}{\mathcal{L}}

\newcommand{\subref}[2]{\ref{#1}\hyperref[#1]{#2}}

\begin{document}

\title{Geometric Dynamics of Magnetization: Electronic Contribution}


\author{Bangguo Xiong} 
\email{bgxiong@physics.utexas.edu} 
\affiliation{Department of Physics, The University of Texas at Austin, Austin, TX 78712, USA} 
\author{Hua Chen} 
\affiliation{Department of Physics, Colorado State University, Fort Collins, CO 80523, USA}
\affiliation{School of Advanced Materials Discovery, Colorado State University, Fort Collins, CO 80523, USA}
\author{Xiao Li}
\affiliation{Department of Physics, The University of Texas at Austin, Austin, TX 78712, USA} 
\author{Qian Niu}
\affiliation{Department of Physics, The University of Texas at Austin, Austin, TX 78712, USA} 
\affiliation{ICQM and CICQM, School of Physics, Peking University, Beijing 100871, China} 

\begin{abstract}
To give a general description of the influences of electric fields or currents on magnetization dynamics, we developed a semiclassical theory for the magnetization implicitly coupled to electronic degrees of freedom. In the absence of electric fields the Bloch electron Hamiltonian changes the Berry curvature, the effective magnetic field, and the damping in the dynamical equation of the magnetization, which we classify into intrinsic and extrinsic effects. Static electric fields modify these as first-order perturbations, using which we were able to give a physically clear interpretation of the current-induced spin-orbit torques. We used a toy model mimicking a ferromagnet-topological-insulator interface to illustrate the various effects, and predicted an anisotropic gyromagnetic ratio and the dynamical stability for an in-plane magnetization. Our formalism can also be applied to the slow dynamics of other order parameters in crystalline solids.
\end{abstract}
\pacs{75.78.-n, 
75.60.Jk, 
75.76.+j 
}
\maketitle 

\noindent{\it Introduction}---Magnetization dynamics is conventionally described by the phenomenological Landau-Lifshitz-Gilbert (LLG) equation, in which the effective magnetic field and the damping factor can be associated with various mechanisms such as dipolar interaction, exchange coupling, electron-hole excitations, etc., through microscopic theories. The recently discovered current-induced spin-orbit torques emerge as current-dependent modifications to the LLG equation, and can be consequently categorized as field-like and damping-like torques \cite{manchon:2008aa,garate:2009aa,Pesin2012,wang2012,Freimuth:2014,KurebayashiH.:2014aa}. In systems with strong spin-orbit coupling and broken inversion symmetry, e.g. GaMnAs, heavy-metal/ferromagnet bilayers and magnetically doped topological insulator heterostructures, magnetization switching using electric current alone through the spin-orbit torque has been achieved experimentally \cite{chernyshov:2009aa, KurebayashiH.:2014aa, Mellnik:2014spin, Fan:2014aa}. Theoretical studies of spin-orbit torques have mostly adopted $s-d$ type couplings between transport electrons and those contributing to magnetization \cite{manchon:2008aa,garate:2009aa,Pesin2012,wang2012,KurebayashiH.:2014aa}, or a self-consistent-field picture based on the spin density functional theory \cite{Freimuth:2014,qian:2002spin}. Then the spin-orbit torques can be understood as the modification to the effective exchange fields proportional to the current-induced spin densities in inversion symmetry breaking systems, known as the Edelstein effect \cite{edelstein:1990spin,garate:2009aa}. However, in general neither the size of the exchange field nor its dependence on order parameter (magnetization) direction is known \textit{a priori} \cite{brataas:2014aa,Hals2013}. It is thus more desirable to develop a theoretical framework that does not explicitly depend on the details of the coupling between transport electrons and the magnetization.

In this Rapid Communication, we provide a semiclassical framework for the dynamics of magnetization implicitly coupled to electronic degrees of freedom, based on the wave-packet method. We found that the Bloch electrons yield a Berry curvature $\Omega_{\bm{m}\bm{m}}$, acting as a magnetic field in the \emph{magnetization space}, while the gradient of the electronic free energy with respect to the magnetization acts as a static \emph{electric field} in the magnetization space, in agreement with previous adiabatic theory of magnetization dynamics \cite{niu:1999adiabatic}. These two fields thus govern the dynamics of magnetization as that of Lorentz force to a charged particle. In addition, we identified an extrinsic contribution to the magnetization dynamics, corresponding to the Gilbert damping in the LLG equation, which is not included in the adiabatic theory. A static electric field enters the magnetization equation of motion by modifying the Berry curvature $\Omega_{\bm{m}\bm{m}}$, the effective field, and the damping factor as a first-order perturbation. In particular, the modification to the effective field includes a part proportional to the Berry curvature $\Omega_{\bm{mk}}$ and having a geometric nature. We used a simplified model for the ferromagnet-topological-insulator interface to illustrate the various effects, and showed that the gyromagnetic ratio is renormalized anisotropically and that an in-plane magnetization can be dynamically stable under moderate electric fields. 

\noindent{\it Formulation and general results}---We start from a general Hamiltonian of Bloch electrons implicitly depending on the order parameter $\bm m$, $\hat H_e(\bm{q};\bm{m})$, where $\bm{q}$ is the crystal momentum. External electromagnetic fields are described by the scalar and vector potentials $(\phi,\bm{A} )$ that enter the Hamiltonian through minimum coupling ($\hbar = 1$, $e = |e|$), 
 \be
  \hat H=\hat H_e(\bm{q} + e\bm{A};\bm{m})-e  \phi.
 \ee  
Following Ref.~\onlinecite{sundaram:1999aa}, a wave packet is constructed with center position $\bm{x}$ and center physical momentum $\bm{k}$ from the Bloch eigenstates of the local electronic Hamiltonian. The Lagrangian of a single wave packet reads as
\be
\Lagr= \dot{\bm{x}} \cdot  [\bm{k}-e \bm{A}(\bm{x},t)] + \dot{\bm{k}} \cdot \bm A_{\bm{k}} + \dot {\bm{m}} \cdot  \bm A_{\bm{m}} - [\varepsilon-e\phi(\bm{x},t)], \label{eq:lagr}
\ee
with $\bm A_{\bm \lambda} = i \bra {u} {\nabla_{\bm \lambda} u} \rangle$ $(\lambda= \bm{k} \text{ or } \bm{m})$ the Berry connections of the Bloch state $\ket {u}$, and $\varepsilon$ the wave packet energy. For notational simplicity we have dropped the band index. The Lagrangian depends on $(\bm{x},\bm{k})$ of the wave packets and magnetization $\bm{m}$. Thus a set of coupled equations of motion for all three variables can be derived from the Lagrangian principle \cite{niu:2010rmp}:
\bea
&&\dot{\bm{k}} = -e\bm{E}  , \label{eq:DK}  \\
&&\dot{\bm{x}} =\ \  \fracp{\varepsilon}{\bm{k} } + \dot{\bm{k}}\cdot \Omega_{\bm{k} \bm{k} }+\dot {\bm{m}} \cdot \Omega_{\bm{m}\bm{k} }, \label{eq:DX} \\
&&\int [d\bm{k}] f \left (\dot {\bm{m}} \cdot \Omega_{\bm{m}\bm{m}}+ \dot{\bm{k}} \cdot \Omega_{\bm{k}\bm{m}}+\fracp{\varepsilon}{\bm{m}}\right )=0,  \label{eq:DM}
\eea
where the Berry curvatures 
$
\Omega_{\lambda_i\lambda_j} = -2 {\rm Im} \langle {\partial u /\partial \lambda_i } | {\partial u/ \partial \lambda_j} \rangle
$, $\bm \lambda$ = $\bm{k}$ or $\bm{m}$. Eq.~\ref{eq:DM} is obtained by summing over all occupied states, and $f$ is the distribution function for the electrons. Note the magnetization dynamics enters the electron equations of motion through $\Omega_{\bm{k} \bm{k} }$ in Eq.~\ref{eq:DX}, and the terms in the square brackets of Eq.~\ref{eq:DM} can be viewed as conjugates of the right hand side of Eq.~(\ref{eq:DX}), by interchanging $\bm{k}$ and $ \bm{m}$. This is a manifestation of the reciprocity between charge pumping due to magnetization precession and electric-current-induced spin-orbit torque.

The nonequilibrium response of the electrons to an external electric field and/or a dynamical $\bm m$ is accounted for using the semiclassical Boltzmann equation, according to which the deviation of the distribution function from the equilibrium Fermi-Dirac distribution $f_0[\varepsilon(\bm{k},\bm{m})]$ is
\bea
\delta f  
= - \tau \fracp{f_0}{\varepsilon}\left (\dot{\bm{k}} \cdot \fracp{\varepsilon}{\bm{k}} + \dot {\bm{m}} \cdot \fracp{\varepsilon}{\bm{m}} \right),  \label{eq:DF}
\eea
where we have assumed a grand canonical ensemble with fixed temperature and chemical potential. $\tau$ is the relaxation time which we take as a constant for simplicity. Generalization to including more specific scattering mechanisms is straightforward but involved, and does not necessarily provide additional insight on the main issues considered in this work. 

The equations (\ref{eq:DK}-\ref{eq:DF}) complete our semiclassical description of coupled magnetization and electron dynamics in the presence of external electric fields, though they can be easily extended to including magnetic fields and other perturbations. 

In the absence of electric fields, $\dot{\bm k} = 0$, and we can obtain from Eq.~(\ref{eq:DF}) and Eq.~(\ref{eq:DM}) the following equations of motion of the magnetization,
\be
\dot{\bm{m}}\cdot (\bar \Omega_{\bm{m}\bm{m}}+\eta_{\bm{m}\bm{m}}) - \bm{H}  = 0, \label{eq:memo}
\ee  
in getting which we have ignored higher order $\dot{\bm{m}}^2$ terms by assuming that the magnetization dynamics is slow compared to typical electronic time scales. The Berry curvature $\bar \Omega$, the damping coefficient $\eta$ and the effective field $\bm{H}$ in the equation above are respectively
\bea
&&\bar \Omega_{\bm{m}\bm{m}} = \int [d\bm{k}] f_0 \Omega_{\bm{m}\bm{m}},  \\
&&\eta_{\bm{m}\bm{m}} =  -\tau \int [d\bm{k}] \fracp{f_0}{\varepsilon} \fracp{\varepsilon}{\bm{m}} \fracp{\varepsilon}{\bm{m}}, \label{eq:FF}  \\
&&\bm{H} = -\fracp{G}{\bm{m}}, \label{eq:SE}
\eea
where $G$ is the free energy of the electron system. For non-interacting electrons $G = -{\beta}^{-1} \int [d\bm{k}] \ln[1+e^{-\beta(\varepsilon-\mu)}]$ for a single band, where $\beta = 1/k_B T$. Interaction effects may be included in $G$ through different levels of approximations, which will also modify the way magnetization appears in $G$. At this point we will leave $G$ as a general electron free energy depending on $\bm m$ implicitly.

We only consider the transverse modes ($\dot{\bm m}$ perpendicular to $\bm m$) of the magnetization dynamics in this work, although Eq.~\ref{eq:memo} can be used for the longitudinal mode as well. The magnetization is thus described by the polar angle $\theta$ and the azimuthal angle $\phi$. Eq.~(\ref{eq:memo}) can then be converted to the familiar form of the LLG equation,
\be
\dot{\bm{m}} = - \gamma \bm{m}\times \left({\bm{H} - \eta_{\bm{m}\bm{m}} \cdot \dot{\bm{m}}}\right), \label{eq:dyn}
\ee
where the gyromagnetic ratio $\gamma$ is related to the Berry curvature through
\be
\bar{\bm{\Omega}} =  \bm{m}/ {\gamma m^2},  \label{eq:gamma}
\ee
where $\bar{\Omega}_i = \varepsilon_{ijk} \bar{\Omega}_{jk}/2$ is the vector form of the Berry curvature tensor. Expressions similar to Eq.~(\ref{eq:memo}), but \emph{without the damping term}, have been derived using the adiabatic theory \cite{niu:1998spin}. Since the damping term is explicitly dependent on the relaxation time, which is ultimately due to dissipative microscopic processes such as electron-phonon scattering and electron-impurity scattering, we call it extrinsic contribution to the magnetization dynamics. Note Eq.~\ref{eq:FF} suggests $\eta$ is positive definite, which means it always leads to energy dissipation through Eq.~\ref{eq:dyn}. The remaining terms are intrinsic contributions from the electron degrees of freedom. In particular, from Eq.~(\ref{eq:memo}) one can see that the two intrinsic terms are formally similar to the Lorentz force of a charged particle, with the antisymmetric part of $\Omega_{\bm{m}\bm{m}}$ (or equivalently the vector form $\bm \Omega_{\bm{m}}$) analogous to the magnetic field and $\bm H$ playing the role of the electric field. 

Electric fields enter our formalism through the equation of motion for $\bm k$ [Eq.~(\ref{eq:DK})], which makes the 2nd term in the integrand of Eq.~\ref{eq:DM} nonzero and also contributes to the nonequilibrium distribution function $\delta f$ in  Eq.~(\ref{eq:DF}). After some algebra, we arrive at the same equation as Eq.~(\ref{eq:memo}), but with $\bm H$, $\bar{\Omega}_{\bm m \bm m}$, and $\eta_{\bm m \bm m}$ acquiring the following corrections proportional to the electric field:
 \bea
&&\bm{H}^E=e \bm{E}\cdot  \int [d\bm{k}]\left( \Omega_{\bm{k}\bm{m}}f_0 -\tau\fracp{\varepsilon}{\bm{k}} \fracp{\varepsilon}{\bm{m}} \fracp{f_0}{\varepsilon} \right ),\label{eq:HE}\\
&&\bar{\Omega}_{m_im_j}^E =  e\tau \bm E \cdot  \\\nonumber
&&\;\; \int [d\bm{k}] \left [  \fracp{\varepsilon}{\bm{k}}   \Omega_{m_im_j} - \left ( \Omega_{\bm{k}m_i} \fracp{\varepsilon}{m_j}\right)_{\rm A}\right ]\fracp{f_0}{\varepsilon}, \label{eq:domegaE}\\
&& \eta_{m_im_j}^E =  e\tau \bm E \cdot \int [d\bm{k}] \left ( \Omega_{\bm{k}m_i} \fracp{\varepsilon}{m_j} \right)_{\rm S}\fracp{f_0}{\varepsilon} \label{eq:detaE}
 \eea
where subscript S (A) means the part of $ \Omega_{\bm{k}m_i} \fracp{\varepsilon}{m_j}$ that is symmetric (antisymmetric) under $i\leftrightarrow j$. We next discuss the physical meanings of these results in detail.

For the correction to the effective field, $\bm H^E$, the first term in Eq.~\ref{eq:HE} has a geometric nature and is an intrinsic contribution from the Fermi sea electrons. It is of $\Omega_{\bm{m}t}$ type, where the time variation is due to the momentum change of a single wave packet driven by $\bm E$: $\partial_t = \dot{\bm{k}}\cdot \partial_{\bm{k}}  = -e\bm{E} \cdot \partial_{\bm{k}}$. We note there is a nice identity connecting $\Omega_{\bm{m}t}$ and the ``magnetic field" in magnetization space $\bm \Omega_{\bm m}$:
\begin{eqnarray}
\partial_t \bm{{\Omega}}_m + \nabla_{\bm{m}} \times \Omega_{\bm{m}t}= 0.
\end{eqnarray}
Since $\Omega_{\bm{m}t} = \Omega_{\bm{m}\bm k}\cdot (-e\bm E)$ is a correction to the static effective \emph{electric field} $\bm H$ (Eq.~\ref{eq:SE}) in the magnetization space, above equation is a magnetic analog of the Faraday's law for charged particles. The 2nd term in Eq.~\ref{eq:HE} is extrinsic since it is proportional to $\tau$, and does not have an electromagnetism analog.

$\bm H^E$ also provides new insights on the charge pumping effect of a nonzero $\dot{\bm m}$ \cite{Freimuth:2015}. Since $P\equiv \bm H^E\cdot \dot{\bm m}$ has the meaning of power density and is proportional to $\bm E$, there is an electric current induced by $\dot{\bm m}$ as $\bm j_p = \partial(\bm H^E\cdot \dot{\bm m})/\partial \bm E$. The change of the polarization density (``pumping") after $\bm m$ completes a closed path in its configuration space is obtained by integrating $\bm j_p$ over this period. A finite charge pumping thus corresponds to a nonzero work density, and is related to the curl of $\bm H^E$ in the magnetization space through
\begin{eqnarray}\label{eq:Work}
W = \oint \bm j_p\cdot \bm E dt = \oint \bm H^E\cdot d\bm m \\\nonumber
= \iint \nabla_{\bm m}\times \bm H^E \cdot d\bm \sigma_{\bm m},
\end{eqnarray}
where we have used the Stokes theorem, and $d\bm \sigma_{\bm m}$ is the infinitesimal area in the magnetization space. Thus in order to have finite charge pumping, $\bm H^E$ must not be conservative, i.e., it cannot be written as a gradient of certain scalar free energy.
 
We now move on to $\bar{\Omega}_{\bm m \bm m}^E$ and $\eta_{\bm m\bm m}^E$, which are all Fermi surface contributions due to the non-equilibrium part of the distribution function $\delta f$. They are important in magnetic metals and should be discussed on an equal footing as $\bm{H}^E$ for current-induced effects on magnetization dynamics. In the form of Eq.~\ref{eq:dyn}, $\bar{\Omega}_{\bm m \bm m}^E$ renormalizes the gyromagnetic ratio as $\gamma^{\prime} = \gamma/(1+\gamma/\gamma^E)$, where $\gamma^E \equiv 1/\bm m \cdot \bar{\bm \Omega}^E$, while $\eta_{\bm m\bm m}^E$ modifies the damping tensor as $\eta^\prime = \eta + \eta^E$. It is interesting to note that $\eta^E$ does not have to be positive definite. A negative definite total damping will make the free energy minima dynamically unstable while the maxima dynamically stable. Thus in addition to the potential of switching the magnetization between different easy directions, a suitably chosen electric field can in principle switch the magnetization between easy and hard directions, which provides a new mechanism (though volatile) for current driven reading and writing processes in magnetic memory devices.

Before ending this section, we translate our results Eq.~(\ref{eq:HE}-\ref{eq:detaE}) into the commonly used spin-orbit torque language. For small electric fields they can be converted to additional terms added to the right hand side of the LLG equation Eq.~\ref{eq:dyn}:
\be
\dot{\bm{m}} = - \gamma \bm{m}\times \left({\bm{H} - \eta_{\bm{m}\bm{m}} \cdot \dot{\bm{m}}}\right) - \gamma \bm{\tau}_{so}, \label{eq:dynnew}
\ee 
where $\bm{\tau}_{so} = \bm{\tau}_{so}^H +\bm{\tau}_{so}^\gamma +\bm{\tau}_{so}^\eta $ with the separate terms being
\begin{eqnarray}
&&\bm{\tau}_{so}^{H}=  \bm{m} \times \bm{H}^E,\label{eq:taoH}\\
&&\bm{\tau}_{so}^{\gamma}=-\gamma/\gamma^E\bm{m} \times (\bm{H}+\eta\gamma \bm{m}\times \bm{H}),\label{eq:taogamma}\\
&&\bm{\tau}_{so}^{\eta}=\gamma \eta^E \bm{m}\times(\bm{m}\times \bm{H}).\label{eq:taoeta}
\end{eqnarray}
For the special $s-d$ type coupling, $H^E$ is proportional to the spin density response to electric fields since $\partial \hat{H} /\partial \bm{m} \sim \bm{s}$, in agreement with previous studies \cite{edelstein:1990spin, Mellnik:2014spin, garate:2009aa,Freimuth:2014}, though our formalism is not limited to this coupling form. Morever, there are additional torques $\bm{\tau}_{so}^{\gamma}$ and $\bm{\tau}_{so}^{\eta}$ that cannot be directly explained using spin density response to electric fields. They can, however, always be classified into either field-like or damping-like torques depending on whether there is a sign change upon $\bm m\rightarrow -\bm m$.
 
\begin{figure}[!htb]
\begin{center}
\includegraphics[width=2.6 in]{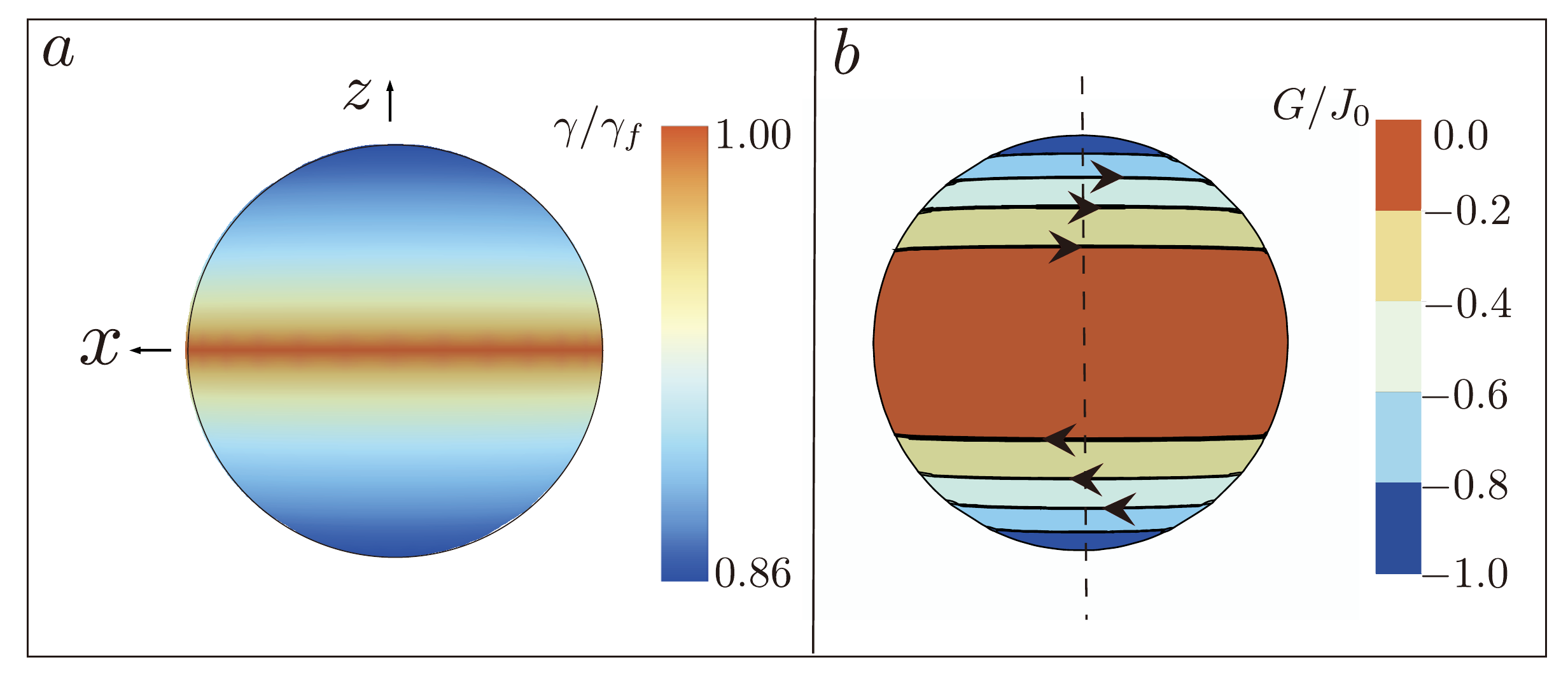}
\end{center} 
\caption{(Color online)  (a) Renormalized gyromagnetic ratio $\gamma$ through coupling to the topological surface states. (b) Contour plot of free energy $G^s$ in the absence of electric fields. The arrows indicate the directions of magnetization motion. Parameters: $\gamma_f = 2ma^2/\hbar, \alpha = 1$}
\label{fig:1}
\end{figure}

\noindent{\it Model example}---As a concrete example, we consider a 2D toy model that can be used to describe the interface between a ferromagnetic insulator and a 3D topological insulator (TI) \cite{Garate:2010, Yokoyama:2010, tserkovnyak:2012aa}:
\be
\hat H({\bm{m} }) = \hbar v  (-k_y \sigma_x + k_x\sigma_y) + J {\bm{m} }\cdot \bm{\sigma},  \label{eq:Hdirac}
\ee
where $\bm m$ is the 2D magnetization of the ferromagnet, $\bm \sigma$ is the Pauli matrix vector for the spin operators, $v$ is the Fermi velocity of the Dirac surface electrons of the TI, and $J$ is the exchange coupling strength between $\bm m$ and $\bm \sigma$. The exchange coupling opens a gap proportional to the $z$ component of $\bm m$. We consider zero temperature and set the chemical potential $\mu=0$. The Berry curvature of the lower band is calculated similarly as the $\vec k\cdot \vec \sigma$ model \cite{niu:2010rmp}
\be
\bar \Omega^s_{\theta\phi} = \frac{\alpha^2|\sin2\theta|}{8\pi a^2} {\rm sgn}(\alpha),
\ee
where $\alpha = Jma/\hbar v$ is the exchange energy measured in typical scales of the kinetic energy $\epsilon_0 = \hbar v /a$ ($a$ is the lattice constant). Using relation Eq.~(\ref{eq:gamma}), the Berry curvature gives an anisotropic gyromagnetic ratio 
\begin{eqnarray}
\gamma_s(\theta) = \frac{4\pi ma^2}{\hbar \alpha^2|\cos\theta|} {\rm sgn}(\alpha).
\end{eqnarray}
We should note that the ferromagnet by itself has a gyromagnetic ratio, denoted as $\gamma_f$, and the overall gyromagnetic ratio $\gamma$ is corrected as 
\be
\gamma^{-1} ={\gamma_f}^{-1} + {\gamma_s}^{-1},
\ee 
or equivalently
\be
\gamma = \gamma_f \cdot \frac{1}{1+\gamma_f/\gamma_s(\theta)}.
\ee
The variation of $\gamma$ for $\bm m$ moving across the Bloch sphere is shown in Fig.~\subref{fig:1}{(a)}. On the equator ($\theta = \pi/2$), $\gamma  = \gamma_f$; at the north and south poles, $\gamma = \gamma_f/(1+\gamma_f \hbar \alpha^2 /4\pi ma^2{\rm sgn}(\alpha))$. This angular dependence of gyromagnetic ratio should be able to be detected by ferromagnetic resonance experiments in such systems. 

The free energy density at zero temperature is calculated by integrating the energy of the lower bands. Ignoring a constant term, we get
\be
G^s= - J_0 m_z^2
\ee
where $J_0= {\epsilon_0 k_c  \alpha^2}/{4\pi m^2a} $ and $k_c$ is the momentum cutoff. $G^s$ has two minima at the north and south poles, as shown in Fig.~\subref{fig:1}{(b)}. Thus the surface states provide a perpendicular magnetic anisotropy for the ferromagnet. For simplicity we ignored the magnetic anisotropy energy of the ferromagnet itself. For nonzero $m_z$ there is no contribution from the surface state electrons to $\eta$ because of the finite gap, and if the intrinsic damping of the ferromagnet is ignorable the magnetization should move along equal-energy lines without driving forces, along the directions determined by $- \gamma \bm{m} \times \bm{H}$ [Eq.~(\ref{eq:dyn})], as illustrated in Fig.~\subref{fig:1}{(b)}.

\begin{figure}[!htb]
\begin{center}
	\includegraphics[width=2.6 in]{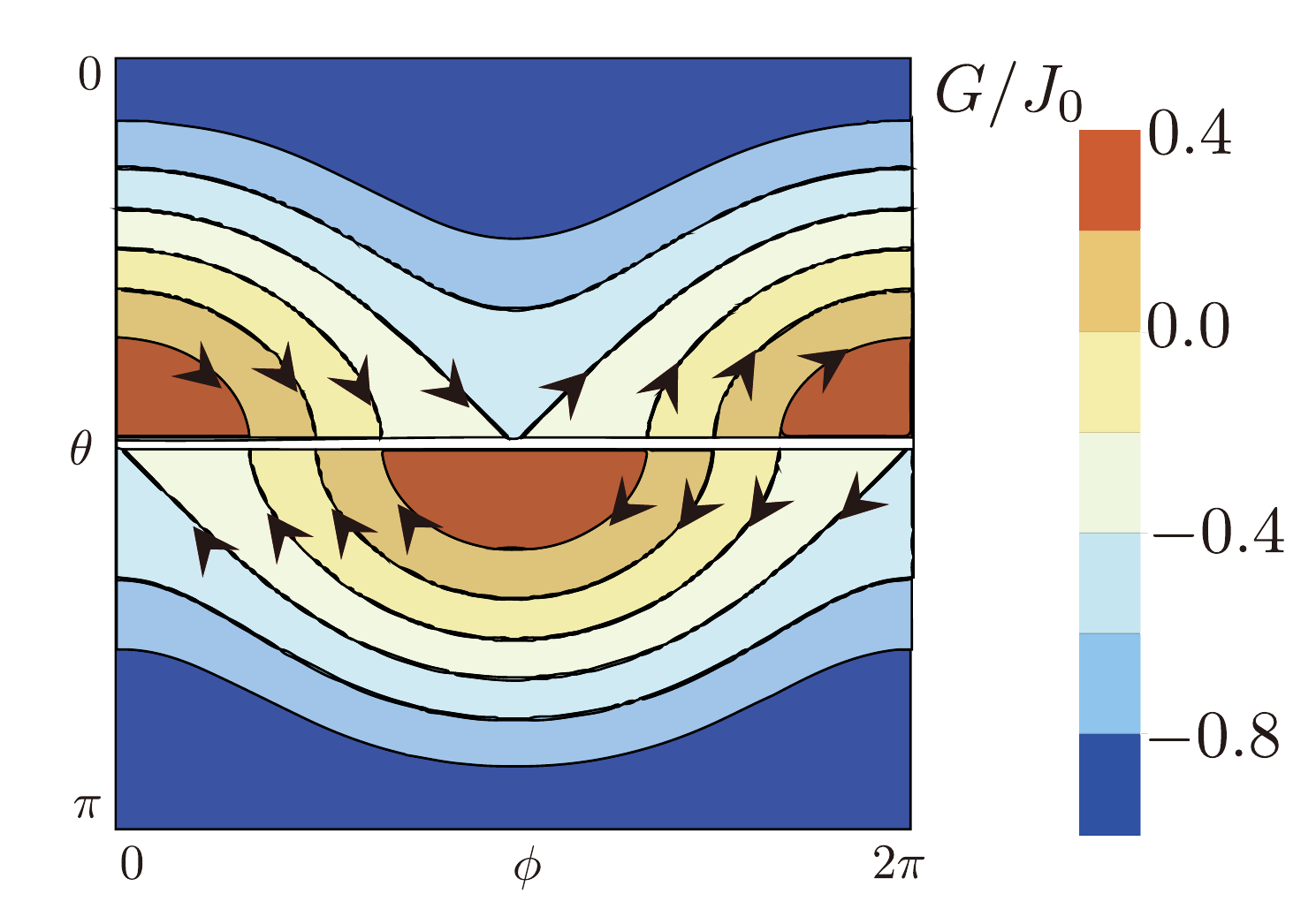}
\end{center} 
\caption{(Color online)    Contour plot of free energies $G_N$ and $G_S$ in the presence of an electric field along $\hat{x}$. Parameters: $\gamma_f = 2ma^2/\hbar, \alpha = 1$, $eE|\alpha|/4\pi J_0 = 0.4$.}
\label{fig:2}
\end{figure}

We now consider the effect of an electric field along $x$ direction on the magnetization dynamics. For nonzero $m_z$ all Fermi surface contributions in Eqs.~(\ref{eq:HE}-\ref{eq:detaE}) are zero, and the only finite term is the Fermi sea contribution in $\bm{H}^E$:
\be
\bm{H}^E = -\frac{eE|\alpha|}{4\pi ma} {\rm sgn}(m_z) \hat{x}.
\ee 
It has constant magnitude but opposite directions depending on the sign of $m_z$. The curl of $\bm{H}^E$ is thus zero everywhere except on the equator, which also means nonzero charge is pumped by magnetization dynamics when the precession axis is in plane \cite{Ueda:2012}. Based on our discussion in the previous section we can only define free energy functions separately for the north (N) and the south (S) hemispheres as $G_N$ and $G_S$ but not globally:
\begin{subequations}
\bea
G_{N} = -J_0m_z^2 + \frac{eE|\alpha|}{4\pi ma} m_x,  \\ 
G_{S}=  -J_0m_z^2 - \frac{eE|\alpha|}{4\pi ma} m_x.
\eea
\end{subequations}
On each hemisphere, the 2nd term in the free energy implies a magnetization-dependent polarization, which will be interesting to detect experimentally. Moreover, since $G_{N}-G_{S} \propto m_x$, they cannot be connected by a constant energy shift across the equator. The electric field thus shifts the two free energy minima at the north and the south poles in opposite directions, and distorts the equal energy lines in the vertical direction, as shown in Fig. \subref{fig:2}. In addition, the opposite signs of $G_{N}$ and $G_{S}$ very close to the equator make half of the equator dynamically stable, as can be seen from the arrows pointing to the equator from both above and below in Fig. \subref{fig:2}. Specifically, if we still assume a vanishing intrinsic damping of the ferromagnet, when the magnetization is very close to the equator with $\phi \in (\pi,2\pi)$, or more generally when it is between the two critical trajectories determined by $G_{N/S}= -{eE|\alpha|}/{4\pi a}$, it will follow the equal energy lines and end up on the half equator with $\phi \in (0,\pi)$. Conversely, for a magnetization outside of the region between the two critical trajectories, i.e., $G_{N/S} < -{eE|\alpha|}/{4\pi a}$, it will keep precessing around one of the free energy minima. When there is a small damping, the size of the attraction area around the half equator reduces because energy is dissipated during evolution. 

In the limiting case of strong electric fields ${eE|\alpha|}/{4\pi a} > 2J_0 m^2$, the critical trajectories disappear on the Bloch sphere and the magnetization will always evolve to the stable half equator. Since without the magnetic field the magnetization has a perpendicular anisotropy due to the topological surface states, electric fields can lead to dynamical switching between easy (out-of-plane) and hard (in-plane) directions. This mechanism is unique to the FM/TI system and is independent of the easy-hard-axes switching due to a negative-definite damping tensor discussed in the last section.  

Since the electric field enters our formalism only through its modification on momentum [Eq.~(\ref{eq:DK})], our theory can be straightforwardly generalized to other time-varying perturbations that influence wave-packet dynamics in similar ways, which will give both Fermi-surface contributions and Fermi-sea contributions through the Berry curvature $\Omega_{\bm{m}t}$. For example, a potential application is the magnetization dynamics driven by sound wave \cite{scherbakov:2010aa,weiler:2011aa}.  Separately, our formalism can be applied to the slow dynamics of other order parameters in crystalline solids, and to its dependence on electromagnetic fields through the electron degrees of freedom.
 

We acknowledge useful discussions with A. H. MacDonald, R. Cheng, Y. Gao, H. Zhou. This work is supported by National Basic Research Program of China (Grant No. 2013CB921900), DOE (DE-FG03-02ER45958, Division of Materials Science and Engineering), NSF (EFMA-1641101), and the Welch Foundation (F-1255). 

\bibliographystyle{apsrev4-1}    
\balance
\bibliography{MDreference}
\end{document}